\documentclass[aps,showpacs,twocolumn,amsmath,amssymb]{revtex4}
\usepackage{amsmath,amssymb}
\usepackage{graphics,epsfig}
\usepackage{graphicx}
\usepackage{color}

\def\build#1_#2^#3{\mathrel{\mathop{\kern 0pt#1}\limits_{#2}^{#3}}}

 \newcommand{\vct}[1]{{\mbox {\boldmath $#1$}}}

\begin{document}

\title{Local and nonlocal pressure Hessian effects in real and synthetic fluid turbulence}

\author{Laurent Chevillard}
\author{Emmanuel L\'ev\^eque}
\author{Francesco Taddia}
\affiliation{Laboratoire de Physique de l'\'Ecole Normale Sup\'erieure
de Lyon, CNRS, Universit\'e de Lyon, 46 all\'ee d'Italie F-69007 Lyon,
France}
\author{Charles Meneveau}
\author{Huidan Yu}
\affiliation{Department of Mechanical Engineering and Center
for Environmental and Applied Fluid Mechanics, The Johns Hopkins
University, 3400 N. Charles Street, Baltimore, MD 21218,
USA}
\author{Carlos Rosales}
\affiliation{Department of Mechanical Engineering, Technical University Federico Santa Mar\'ia,
Av. Espa\~{n}a 1680, Valparaiso, Chile}

\begin{abstract}
The Lagrangian dynamics of the velocity gradient tensor \textbf{A} in isotropic and homogeneous turbulence depend 
on the joint action of the self-streching term and the pressure Hessian.
Existing closures for pressure effects in terms of  \textbf{A} are unable to reproduce one important statistical 
role played by the anisotropic part of the pressure Hessian, namely the redistribution of the probabilities towards enstrophy production dominated regions. 
As a step towards elucidating the required properties of closures, we study several synthetic velocity fields and how well
they reproduce anisotropic pressure effects. It is found that synthetic (i) Gaussian, (ii) Multifractal and (iii) Minimal Turnover Lagrangian Map (MTLM) incompressible 
velocity fields reproduce many features of real pressure fields that are obtained from numerical simulations of the Navier Stokes equations, including the
redistribution towards enstrophy-production regions.  The synthetic fields include both spatially local, and nonlocal, anisotropic  pressure effects.  However, we show that
the local effects appear to be the most important ones: by assuming that the pressure Hessian is local in space, an expression in terms of the Hessian of the 
second invariant $Q$ of the velocity gradient tensor can be obtained. This term is found to be well correlated with the true pressure 
Hessian both in terms of eigenvalue magnitudes and eigenvector alignments.
\end{abstract}


\maketitle

\section{Introduction}
\label{sec-intro} 

The study of the velocity gradient tensor in fully developed turbulence has 
lead to interesting findings and has contributed to improved understanding of many statistical and geometrical properties of turbulent flows. In particular, recent progress has been made in the study of the Lagrangian dynamics and modeling of the velocity gradient tensor (see Ref.  \cite{Men11} for an overview of the subject). This tensor is given by $A_{ij} = \partial u_i/\partial x_j$, where \textbf{u} is the velocity vector. Taking a spatial gradient of the Navier-Stokes equations,  the following transport equation for \textbf{A} is obtained:
\begin{equation}\label{eq:NSA}
\frac{d \textbf{A}}{dt} = - \textbf{A}^2- \textbf{P}+\nu\Delta  \textbf{A}\mbox{ ,}
\end{equation}
where $d/dt$ stands for the Lagrangian time derivative, $\nu$ is the kinematic viscosity and $P_{ij} = {\partial^2p}/{\partial x_i \partial x_j}$ is the pressure Hessian. 
The first term $-\textbf{A}^2$ is the self-stretching term. The restricted Euler (RE) approximation, which assumes 
an isotropic pressure Hessian $P_{ij} = -\mbox{tr}( \textbf{A}^2) \delta_{ij}/3$  and neglects viscous effects, 
leads to an autonomous set of coupled ordinary differential equations \cite{Cantwell92}. 
The intrinsic dynamics of the RE system leads to a finite time divergence of the components of  \textbf{A} during which the vorticity 
$\vct{\omega} = \vct{\nabla}\wedge \textbf{u}$ gets aligned with the eigenvector of the rate of strain, $\textbf{S} = (\textbf{A}+\textbf{A}^\top)/2$, associated with the intermediate eigenvalue, as often observed in real turbulence \cite{TsinoBook,Wal09,Men11}. To prevent the development of unphysical finite time singularities, 
both the anisotropic part of the pressure Hessian and the viscous diffusion term 
have to be modeled. This was the subject of former works \cite{ChePum99,JeoGir03,CheMen06}. 
In particular, closures were proposed in Ref. \cite{CheMen06} for $\textbf{P}$ and $\nu\Delta  \textbf{A}$ 
in terms of the local value of  \textbf{A}. The local closures of Ref. \cite{CheMen06}, 
when inserted into the dynamics generated by Eq. (\ref{eq:NSA}) under the action of a stochastic forcing term, 
lead to stationary statistics of \textbf{A} along Lagrangian trajectories which compare well with those obtained from direct 
numerical simulations (DNS) of the Navier-Stokes equations at moderate Reynolds numbers \cite{CheMen08}. 
At higher Reynolds numbers, predictions of the stochastic model proposed in Refs. \cite{CheMen06,CheMen08} turn out to 
become unrealistic, mainly because of the weakness of the closure for the anisotropic part of the pressure Hessian. 

Indeed, the pressure Hessian is related to
the spatial distribution of the velocity gradient using singular
integral operators \cite{Ohk93,Con94,OhkKis95,MajBer02,Ohk08} according to:
\begin{equation}\label{eq:SOPH}
\frac{\partial^2 p}{\partial x_i\partial x_j} =
-\mbox{tr}(\textbf{A}^2)\frac{\delta_{ij}}{3}-\mbox{P.V.}\int
k_{ij}(\textbf{x}-\textbf{y})
\mbox{tr}(\textbf{A}^2)(\textbf{y})d\textbf{y}.
\end{equation}
Above,  the integral is understood as a Cauchy principal value
(P.V.) and $k_{ij}$ is the Hessian of the Green's
function for the Laplacian operator, namely
\begin{equation}
k_{ij}(\textbf{x}) = \frac{\partial ^2}{\partial x_i\partial x_j}
\frac{1}{4\pi|\textbf{x}|} =
\frac{|\textbf{x}|^2\delta_{ij}-3x_ix_j}{4\pi|\textbf{x}|^5}
\mbox{ .}
\end{equation}
One can see from Eq. (\ref{eq:SOPH}) that only the isotropic part
of the pressure Hessian is purely local (the first term on the right-hand side (RHS) of Eq. (\ref{eq:SOPH})). All the nonlocal effects of pressure
Hessian enter through the anisotropic part (or deviatoric part
corresponding to the second term in the RHS of Eq. (\ref{eq:SOPH})).
Hence, in this view, the RE approximation can be understood as the
neglect of all the nonlocal effects implied by the
incompressibility condition (or pressure field).

In order to quantify the precise action of pressure in numerical turbulent flows, it was proposed, in 
Ref. \cite{CheMen08}, to study the probability current associated with pressure in the plane spanned by the two highly relevant invariants of $\textbf{A}$, $R$ and $Q$ (the so-called $RQ$-plane).  One of these invariants,
defined as,
\begin{equation}\label{eq:DefQ}
Q = -\frac{1}{2}\mbox{tr}(\textbf{A}^2)=\frac{1}{4}|\vct{\omega}|^2-\frac{1}{2}
\mbox{tr}(\textbf{S}^2)
\end{equation} 
quantifies the net balance, or competition, between enstrophy and dissipation. The other important invariant, defined as 
\begin{equation}\label{eq:DefR}
R = -\frac{1}{3}\mbox{tr}(\textbf{A}^3)=-\frac{1}{4}\omega_iS_{ij}\omega_j-\frac{1}{3}
\mbox{tr}(\textbf{S}^3),
\end{equation} 
quantifies the competition between enstrophy production and strain skewness 
(i.e. dissipation production). 
As it will be recalled in the following, in terms of the velocity gradient evolution in statistically stationary turbulence, pressure has two important roles. 
First of all, pressure counteracts the development of the singularity implied by the self-stretching 
term. This feature is found to be well reproduced by existing closures  \cite{CheMen08}. 
The other important pressure action  is the redistribution of probabilities towards enstrophy production dominated regions 
(i.e. towards $R<0$). 
This is not reproduced well by existing closures. As discussed in  Refs. \cite{CheMen08,GibHol07}, a related  deficiency 
of the closures is that they all predict that
the pressure Hessian is  proportional to $Q$. In the $R-Q$ plane dynamics, this implies that when $Q=0$,  the effect of pressure Hessian also vanishes. For real turbulence,
there is no such vanishing of pressure Hessian effects when $Q=0$ \cite{CheMen08}. 

In this article, we investigate whether these particular features of pressure 
(redistribution of probabilities towards enstrophy production, and non-vanishing action even when $Q=0$) is 
inherent to true Navier-Stokes turbulence, or can also be observed in various approximations, namely  
synthetic turbulent velocity fields.  Various types of synthetic fields are considered. 
The first type of synthetic field considered is Gaussian fields obtained by superposing random-phase Fourier modes with 
prescribed spectra. The second type is  called `multifractal' \cite{CheRob10}, and consists of a Gaussian field  whose 
vorticity field is amplified by means of the `Fluid Deformation Closure' and made consistent with multifractality's 
long-range correlations in physical space \cite{Fri95}. 
The third type of synthetic field is generated using the Lagrangian mapping technique \cite{RosMen06,RosMen08}.  
It also relies on random-phase Gaussian fields but then applies a multi-scale deformation of  fluid particles  
using a simple Lagrangian mapping. For each of these synthetic velocity fields, a pressure field is obtained numerically by means 
of the pressure Poisson equation. As will be seen, unlike the local closures discussed above, these synthetic fields 
reproduce many correct features of the pressure Hessian. In particular, they will be shown to reproduce the redistribution 
of probability towards enstrophy production, as well as displaying non-vanishing action, even when $Q=0$.

The second part of the paper studies to what degree spatial locality is important in determining these properties of the
anisotropic part of the pressure Hessian. As can be seen from the expression for the pressure Hessian (Eq. (\ref{eq:SOPH})), 
the anisotropic part of the pressure Hessian
is also the part that is spatially nonlocal, i.e. the part that requires knowledge of $\mbox{tr}(\textbf{A}^2)$ at  
positions ${\bf y} \neq {\bf x}$. Arguably, the more non-local effects are important, the more challenging it is to formulate  
closures in terms of local quantities.  
In order to examine the degree of locality, in the second part of this paper we decompose  the space integration  in 
Eq. (\ref{eq:SOPH}) into two parts, a local part given by the integration over a small  ball of radius given by the 
Kolmogorov length scale $\eta_K$, and the remainder being the `nonlocal' portion. We will show that neglecting the second 
non-local contribution leads to an expression that models the anisotropic part of the  pressure Hessian
in terms of the Hessian of the invariant $Q$. Using DNS data, this expression is compared with the true pressure Hessian.

\section{Pressure Hessian from DNS and synthetic velocity fields}

\subsection{Probability current in the $RQ$-plane}

\subsubsection{Definition of the probability current}

We follow the approach used in Refs. \cite{BosTao02,CheMen08}, based on a
Fokker-Planck equation for the dynamics of $R$ and $Q$. To
summarize the approach, we remark that it can be shown that the time evolution along a Lagrangian trajectory of the non-dimensional invariants $R^*=R/\sigma^{3}$ and $Q^*=Q/\sigma^2$ is given by 
\begin{equation}\label{eq:DynQ}
\frac{dQ^*}{dt^*} = -3R^*-\frac{1}{\sigma^3}A_{ik}H_{ki}^p
-\frac{1}{\sigma^3}A_{ik}H_{ki}^{\nu} ~~\mbox{ and}
\end{equation}
\begin{equation}\label{eq:DynR}
\frac{dR^*}{dt^*} =
\frac{2}{3}\left(Q^*\right)^2-\frac{1}{\sigma^4}A_{ik}A_{kl}H_{li}^p
-\frac{1}{\sigma^4}A_{ik}A_{kl}H_{li}^{\nu} \mbox{ ,}
\end{equation}
where $\sigma^2 = \langle\mbox{tr}(\textbf{S}^2) \rangle$ is the 
strain variance, and $t^*=\sigma t$ is the non-dimensional time. 
Also, $\textbf{H}^p$ stands for (minus) the deviatoric part of the
pressure Hessian, i.e.
\begin{equation}\label{eq:DeviatoricHP}
H_{ij}^p =-\left( \frac{\partial^2p}{\partial x_i\partial x_j
}-\frac{\delta_{ij}}{3}\frac{\partial^2p}{\partial x_k\partial x_k
}\right)\mbox{ ,}
\end{equation}
and $\textbf{H}^\nu = \nu \nabla^2\textbf{A}$ is the viscous term
(recall that in the RE approximation,
$\textbf{H}^p=\textbf{H}^\nu=0$). The Fokker-Planck equation
describing the time evolution of the joint density $\mathcal
P(Q^*,R^*)$ may be written as:
\begin{equation}\label{eq:FokkerRQ}
\frac{\partial \mathcal P}{\partial t^*} + \begin{pmatrix}
  \frac{\partial }{\partial Q^*} \\
  \frac{\partial }{\partial R^*}
\end{pmatrix}.\vct{\mathcal W}=0\mbox{ ,}
\end{equation}
where the divergence of the \textit{probability current}
$\vct{\mathcal W}$ controls the time variations of the joint
probability density $\mathcal P$. The probability current can be
written in terms of conditional averages as
\begin{equation}\label{eq:WTotal}
\vct{\mathcal W} =\left\langle
\begin{pmatrix}
  \frac{dQ^*}{dt^*} \\
  \frac{dR^*}{dt^*}
\end{pmatrix}\bigg | Q^*,R^*\right\rangle \mathcal P(Q^*,R^*)\mbox{ ,}
\end{equation}
and can be decomposed into $\vct{\mathcal W} = \vct{\mathcal
W}_{RE}+\vct{\mathcal W}_p+\vct{\mathcal W}_\nu$, where the probability currents $\vct{\mathcal
W}_{RE}$, $\vct{\mathcal W}_p$ and $\vct{\mathcal W}_\nu$ are associated, respectively, 
to the effects on the Lagrangian evolution of the invariants $Q$ and $R$ (Eqs. (\ref{eq:DynQ}) and (\ref{eq:DynR})) of the 
restricted Euler term $-\textbf{A}^2$, of (minus) the pressure Hessian $-\textbf{P}$
and of diffusivity $\nu\nabla^2 \textbf{A}$ entering in Eq. (\ref{eq:NSA}). In this article, we will focus on the probability current associated 
with the pressure Hessian $\vct{\mathcal W}_p$. It can be written as  
\begin{equation}\label{eq:WP}
\vct{\mathcal W}_{p} =\left\langle
\begin{pmatrix}
  -A_{ik}H_{ki}^p/\sigma^3 \\
  -A_{ik}A_{kl}H_{li}^p/\sigma^4
\end{pmatrix} \bigg | Q^*,R^*\right\rangle\mathcal P(Q^*,R^*)\mbox{ .}
\end{equation}
More details are provided in Ref. \cite{CheMen08}.

\subsubsection{DNS velocity fields}
\label{DNSvf}
In the following, we will make extensive use of data from  standard direct
numerical simulation (DNS) of the Navier-Stokes equations, for a
Taylor-based Reynolds number of order $\mathcal R_\lambda=145$.
DNS is based on a pseudo-spectral method with 2nd-order accurate Adams-Bashforth time stepping; 
the computation box is cubic (size $2\pi$) with periodic boundary conditions in the three directions and spatial 
resolution $512^3$. Statistical stationarity is maintained by an isotropic external force acting at low wavenumbers in order 
to ensure a constant energy-power supply. It provides,
in the units of the simulation, a constant energy injection rate $\langle \epsilon \rangle = 0.001$. The kinematic viscosity
of the fluid is $\nu=0.000285$. The Kolmogorov's scale is 
$\eta_K =(\nu^3/\langle \epsilon \rangle)^{1/4}=0.0123$ so that $dx/\eta_K\approx 1$, since $dx=2\pi/512$.

\begin{figure*}[t]
\center{\epsfig{file=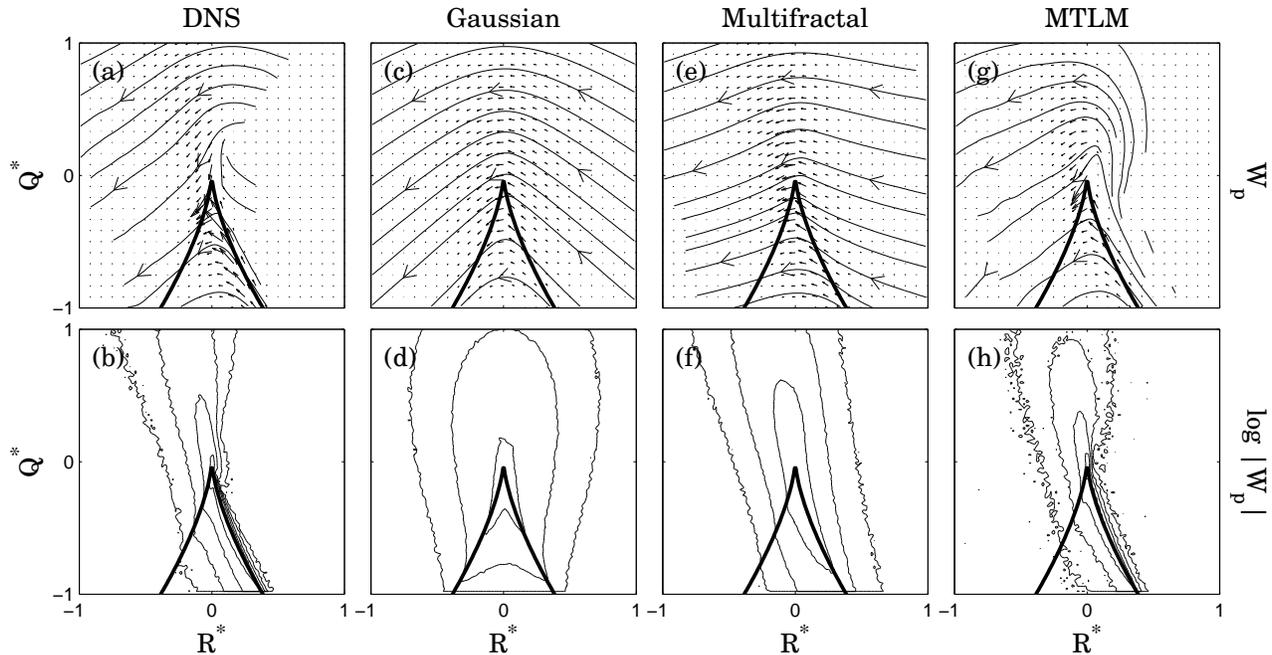,width=17cm}} 
\caption{Probability current $\vct{\mathcal W_p}$ associated with the pressure Hessian --Eq. (\ref{eq:WP})-- for the DNS velocity field and the three synthetic velocity fields --Eqs. (\ref{eq:GaussVect}), (\ref{eq:MultiVel}) 
and (\ref{eq:MTLMVel})-- in the $R^*Q^*$-plane, where $R^* = R/\langle
S_{ij}S_{ij}\rangle^{3/2}$ and $Q^*= Q/\langle
S_{ij}S_{ij}\rangle$. The streamlines and vector plots of $\vct{\mathcal W_p}$ are shown in (a,c,e,g). The iso-probability contours of the magnitude of $\vct{\mathcal W_p}$ are shown in (b,d,f,h). Contours are logarithmically
spaced by factors of 10, starting at 1 for the contour closest to the  origin. The
thick lines represent the zero-discriminant (or Vieillefosse)
line: $\frac{27}{4}R^2+Q^3=0$. } \label{fig:VectAmpRQ}
\end{figure*}

We display in Fig. \ref{fig:VectAmpRQ}(a) the vector plot and streamlines of the probability current 
$\vct{\mathcal W}_p$ associated with the pressure Hessian (Eq. (\ref{eq:WP})) in the $R^*Q^*$-plane, as it was done in 
Ref. \cite{CheMen08}. Three main remarks can be made at this stage: 
(i) first, the pressure Hessian counteracts the development of the finite time singularity along the right tail 
of the Vieillefosse line implied by the RE term, (ii) probabilities are found to be very low in the dissipation production 
dominated region (i.e. $R^*>0$ and above the Vieillefosse tail) meaning that pressure does not play there a significant role, and (iii) pressure 
redistributes the probabilities towards the enstrophy production dominated region (i.e. the flux is directed to the left, towards $R^*<0$). 
As far as the restricted Euler term is concerned, as is well known \cite{Men11,Cantwell92}, the
deterministic $\vct{\mathcal W}_{RE}$ probability current pushes probabilities toward the right tail of the Vieillefosse line 
(data not shown).
This result helps create a picture of the time evolution of velocity gradients along Lagrangian trajectories in stationary flows: 
The RE term ``pushes'' the probabilities towards the right tail towards and along the Vieillefosse line, while the pressure regularizes 
the implied finite time singularity and redistributes the probabilities towards the left part of the plane such that, 
in turn, the RE term can act again, etc. To that picture should be added the viscous diffusion effects, 
$\vct{\mathcal W}_\nu$ pushes the probabilities toward vanishing $R$ and $Q$ not only along the Vieillefosse line but also
everywhere else, and stochastic forcing, such that 
the full probability current $\vct{\mathcal W}$ (Eq. (\ref{eq:WTotal})) is divergence free, in order to ensure stationary 
statistics (Eq. (\ref{eq:FokkerRQ})) \cite{CheMen08}. 

We display also in Fig. \ref{fig:VectAmpRQ}(b) the amplitude of  $\vct{\mathcal W}_p$, using  
logarithmic spacing of iso-probability lines. It can be seen that indeed no current is discernible in the right part of the plane. We remark that It would be interesting 
to quantify whether the effect of anisotropic pressure Hessian  is orthogonal to the $RQ$-plane in this region, when the $RQ$ plane is extended into three-dimensions, as 
proposed and studied in Ref. \cite{LutHol09}.

\subsubsection{Incompressible Gaussian stochastic velocity field with K41 correlation structure}

Let us write a Gaussian homogeneous, isotropic and incompressible vectorial field $\textbf{u}(\textbf{x})$ \cite{RobVar08,TafUns10}, 
having a correlation structure consistent with K41 scalings. It reads, in $d$-dimensions,
\begin{equation}\label{eq:GaussVect}
\textbf{u}_\epsilon(\textbf{x}) = \int_{\mathbb R^d}\varphi_L(\textbf{x}-\textbf{y})
\frac{\textbf{x}-\textbf{y}}{|\textbf{x}-\textbf{y}|_\epsilon^{\frac{d}{2}+\frac{2}{3}}} \wedge d\textbf{W}(\textbf{y})\mbox{ ,}
\end{equation}
where $d\textbf{W}(\textbf{y})=(dW_1(\textbf{y}),dW_2(\textbf{y}),..., dW_d(\textbf{y}))$ is a Gaussian vectorial white noise, and 
$\varphi_L$ is a large-scale cut-off which involves the integral length scale $L$. The deterministic kernel entering into 
Eq. (\ref{eq:GaussVect}) is 
regularized over the small length scale $\epsilon$, namely $|.|_\epsilon = \theta_\epsilon*|.|$ ($*$ stands for the convolution 
product), with a mollifier $\theta_\epsilon(\textbf{x}) = \frac{1}{\epsilon^d}\theta \left( \frac{\textbf{x}}{\epsilon}\right) $ 
and $\int \theta (\textbf{x})d\textbf{x}=1$. It is shown in Ref. \cite{RobVar08} that the velocity $\textbf{u}_\epsilon(\textbf{x})$ 
has a well-defined limit when $\epsilon\rightarrow 0$, denoted by $\textbf{u}(\textbf{x})$, and such that 
$\langle |\textbf{u}(\textbf{x}+\ell\textbf{e})-\textbf{u}(\textbf{x})|^q\rangle \sim C_q (\ell/L) ^{q/3}$ when $\ell\rightarrow 0$,
with $C_q$ a constant independent on the vector $\textbf{e}$.

A Gaussian vectorial field, such as from Eq. (\ref{eq:GaussVect}), is a poor representation of turbulence since it does not reproduce several 
important features such as a mean energy transfer towards small scales (i.e. the skewness phenomenon), the non-Gaussianity of 
velocity increments (i.e. the intermittency phenomenon) and the alignment of vorticity with the intermediate eigenvector of the 
strain rate tensor \cite{TsinoBook,Wal09,Men11}. Nevertheless, it is useful to consider it in the analysis of the statistical quantities in which we are interested, such as the probability current $\vct{\mathcal W}_p$ associated with the pressure Hessian (Eq. (\ref{eq:WP})). In particular, we look at  
which part can be attributed to Gaussian statistics and which part is really linked to turbulence.

The Gaussian velocity field (Eq. (\ref{eq:GaussVect})) is computed in a periodic box in $d=3$ space dimensions, 
using $N=1024^3$ collocation points. The regularizing parameter $\epsilon$ is chosen as $\epsilon=6dx$, where the spatial 
resolution is $dx=1/N$. For the mollifier $\theta$ and the large scale cut-off $\varphi$, we take Gaussian functions. See Ref. \cite{CheRob10} for further numerical details.
The Pressure $p$ is defined via the Poisson equation $\Delta p = -\mbox{tr}(\textbf{A}^2)$, where $\textbf{A}$ is the (Gaussian) 
velocity gradient tensor. 

Figure \ref{fig:VectAmpRQ}(c) shows the vector plot and streamlines of the probability current $\vct{\mathcal W}_p$ 
obtained from the Gaussian velocity field --Eq. (\ref{eq:GaussVect})--. It can be seen that the pressure from the Gaussian field 
does not counteract directly the singularity along the direction of the right tail of the Vieillefosse line, as  is the case in the DNS. 
Also, there is a significantly higher probability current in the right side region (i.e. $R^*>0$ and above the Viellefosse tail) than in the DNS. 
The pressure obtained from the Gaussian field  is only realistic in the enstrophy production dominated region (i.e.  $R^*<0$), where the behavior shows indeed a trend to push  the probability density towards this region. Also, the streamlines cross the $Q=0$ line meaning that the pressure field from the Gaussian 
velocity field does produce non-zero effect even when $Q=0$. 

In Fig. \ref{fig:VectAmpRQ}(d), $\vct{\mathcal W}_p$-amplitude is shown. It can be seen that the amplitude iso-values are symmetric with respect to the $R^*=0$ line.  We are thus led to the conclusion that a Gaussian velocity field, and its associated pressure field, do not make difference between dissipation production dominated regions ($R^*>0$) and enstrophy production dominated regions ($R^*<0$).

\subsubsection{Incompressible Multifractal stochastic velocity field with KO62 statistics}

Based on the recent fluid deformation imposed by the Euler flow \cite{CheMen06}, and further heuristic introduction of the 
multifractal structure of turbulence as observed from extensive empirical data (see e.g. \cite{Fri95}),  Ref. \cite{CheRob10} proposed the following 3D 
random vectorial field:
\begin{equation}\label{eq:MultiVel}
 \textbf{u}_\epsilon(\textbf{x}) = \int_{\mathbb R^3}\varphi_L(\textbf{x}-\textbf{y})
\frac{\textbf{x}-\textbf{y}}{|\textbf{x}-\textbf{y}|_\epsilon^{\frac{3}{2}+\frac{2}{3}}} \wedge e^{\textbf{S}_\epsilon (\textbf{y})}d\textbf{W}(\textbf{y})\mbox{ ,}
\end{equation}
where $\textbf{S}$ is a tensorial Gaussian log-correlated noise of the form
\begin{align}\label{eq:Slog}
\textbf{S}_\epsilon(\textbf{y}) &= \sqrt{\frac{5}{4\pi}}\lambda\int_{|\textbf{y}-\vct{\sigma}|\le L}\left[ \frac{(\textbf{y}-\vct{\sigma})\otimes [(\textbf{y}-\vct{\sigma})\wedge d\textbf{W}(\vct{\sigma})]}{|\textbf{y}-\vct{\sigma}|_\epsilon^{7/2}}\right.\notag \\
& \left.+ \frac{ [(\textbf{y}-\vct{\sigma})\wedge d\textbf{W}(\vct{\sigma})]\otimes(\textbf{y}-\vct{\sigma})}{|\textbf{y}-\vct{\sigma}|_\epsilon^{7/2}}\right]\mbox{ ,}
\end{align}
with $\otimes$ denoting the tensorial product.  
The form of the symmetric matrix $\textbf{S}_\epsilon$ is inspired by the 
recent fluid deformation closure 
experienced by the fluid over short times \cite{CheRob10}, and the exponent $\frac{7}{2}$ has been selected such that the components of 
$\textbf{S}$ are correlated logarithmically in space. A free parameter $\lambda$ enters this construction and governs the level of intermittency 
of the field. We will take in the sequel $\lambda^2=0.025$ in order to be consistent with empirical findings \cite{CheCas06}. 

Generation of the vectorial field $\textbf{u}_\epsilon(\textbf{x})$ can be done accurately and efficiently in periodic boxes 
using up to $1024^3$ collocations points, in a similar way as  done for the Gaussian velocity field (Eq. (\ref{eq:GaussVect})). 
The cost of the computation of the matrix exponential is the limiting numerical step. It is estimated at each point of space using a 
Pad\'e approximant with scaling and squaring (see \cite{CheRob10} for details). 

It has been shown numerically (Ref. \cite{CheRob10}) that the multifractal velocity field --Eq. (\ref{eq:MultiVel})-- gives a realistic representation of instantaneous realizations of velocity fields in fully developed turbulence in the inertial range, in regard to the following properties: (i) Longitudinal $\delta_\ell u$ and transverse velocity increments are intermittent,  $\lambda$ being  the intermittency coefficient, 
(ii) the third-order moment $\langle (\delta_\ell u)^3 \rangle$ is negative and proportional to the scale 
$\ell$. The fact that there is negative skewness $S=\langle (\delta_\ell u)^3 \rangle/\langle (\delta_\ell u)^2 \rangle^{3/2}$ means that  \textbf{u} exhibits a non-vanishing 
mean energy transfer towards the small scales and 
(iii) vorticity  gets preferentially aligned with the eigenvector of the strain-rate tensor corresponding to the intermediate eigenvalue. 

We show in Fig. \ref{fig:VectAmpRQ}(e) the vector plot and streamlines of the probability current $\vct{\mathcal W}_p$ 
obtained from the multifractal velocity field. It can be seen that, in a similar fashion to the Gaussian case, streamlines are roughly symmetric with respect to the $R^*=0$ line. This is not consistent with  DNS in the $R^*>0$ region, but it is still realistic in the $R^*<0$ region. The difference with the Gaussian field is the fact that now the joint 
density of $R^*$ and $Q^*$ is not symmetric with respect to the $R^*=0$ line, showing thus the predominance of the regions for enstrophy-enstrophy 
production (upper-left quadrant, which in turbulent flows is correlated
with vortex stretching) and dissipation-dissipation production (lower-right quadrant, connected with biaxial straining in turbulent flows). This is also the case for the probability current amplitude, as shown in  Fig. \ref{fig:VectAmpRQ}(f). We can see therefore that both the Gaussian and 
Multifractal velocity fields do not reproduce the void in probability in the $R^*>0$ region as observed in DNS, 
but they do reproduce accurately the 
probability current evolutions in the $R^*<0$ regions and the presence of probability flux at $Q=0$.

\subsubsection{MTLM velocity field}
We consider a third case of a synthetic velocity field. The minimal turnover Lagrangian map (MTLM) velocity field is obtained by distorting an initially random solenoidal vector field, $\mathbf{u}_0(\mathbf{x})$, 
over a hierarchy of spatial scales $\{\ell_n = 2^{-n}L, n=1,\ldots,M\}$, where $L$ is of the order of the integral scale, 
and the smallest scale, $\ell_M$, is of the order of Kolmogorov scale. This generates the multiscale recursive sequence
\begin{equation}\label{eq:MTLMVel}
\mathbf{u}_n(\mathbf{x}) = T[\mathbf{u}_{n-1}(\mathbf{x}), \ell_n]\ ; \ \  n=1,\ldots,M	
\end{equation}
whose final step, $\mathbf{u}_M(\mathbf{x})$, is the synthetic velocity field. 
Here, $T[\cdot]$ stands for the distortion operations applied. At each level $n$ in the sequence the velocity is filtered 
at scale $\ell_n$ and decomposed into low-pass and high-pass filtered parts: $\mathbf{u}^<$ and $\mathbf{u}^>$ respectively. 
The $\mathbf{u}^<$ part is deformed by mapping the velocity vectors from their collocation points, $\mathbf{x}$, to new positions 
that fluid particles moving at constant velocity in Lagrangian coordinates would reach: 
$\mathbf{u}^<(\mathbf{X}(t),t) = \mathbf{u}^<(\mathbf{x},0)$, with $\mathbf{X}(t) = \mathbf{x} + t \mathbf{u}^<(\mathbf{x})$. 
The parameter $t$ is taken equal to the eddy-turnover time-scale corresponding to the spatial scale $\ell_n$, computed using standard Kolmogorov scaling. 
New velocity values at the collocation points are obtained by interpolation over nearby velocities that have come into a
neighborhood of radius $\ell_n$ around $\mathbf{x}$ after the mapping. This deformed field $\mathbf{u}^<$ is made solenoidal 
again by projection in Fourier space, and the amplitudes of its Fourier modes are scaled to conform to the target energy spectrum. 
Finally, $\mathbf{u}^<$ is recombined with the $\mathbf{u}^>$ part, which at this stage still remains as a Gaussian field. 

The next generation in the hierarchy will take the complete field $\mathbf{u}$ and will apply the same distortion operations, 
now with the field decomposed at a smaller filtering scale. In this way, the effects are superposed and accumulated over a range 
of spatial scales. Further details and characteristics of these synthetic velocity fields can be found in Refs. \cite{RosMen06,RosMen08}. 
The present MTLM velocity field was generated in a periodic box, using $512^3$ collocation points, with $M=6$ generations in the hierarchy, 
and an energy spectrum corresponding to $\mathcal{R}_{\lambda} \approx 250$.

The results for the probability current $\vct{\mathcal W}_p$ obtained from the MTLM velocity field (Eq. (\ref{eq:MTLMVel})) are 
shown in Figs. \ref{fig:VectAmpRQ}(e-f). When compared with the DNS results (Figs. \ref{fig:VectAmpRQ}(a-b)), we can see 
close agreement of both magnitude and direction of $\vct{\mathcal W}_p$. In particular, the MTLM velocity field reproduces, for 
pressure-related part of the probability current, the void of probability in the $R^*>0$ region. We can conclude that, of the three cases studied, the MTLM velocity field gives the most realistic synthetic turbulence, as far as anisotropic pressure Hessian effects are concerned. But some small difference can be observed in the 
$R^*>0$ region, close to the origin, where the MTLM fields seem to predict a circular motion that is not present in DNS.  

\subsection{Mean pressure Hessian norm conditioned on Q}

\begin{figure}[t]
\center{\epsfig{file=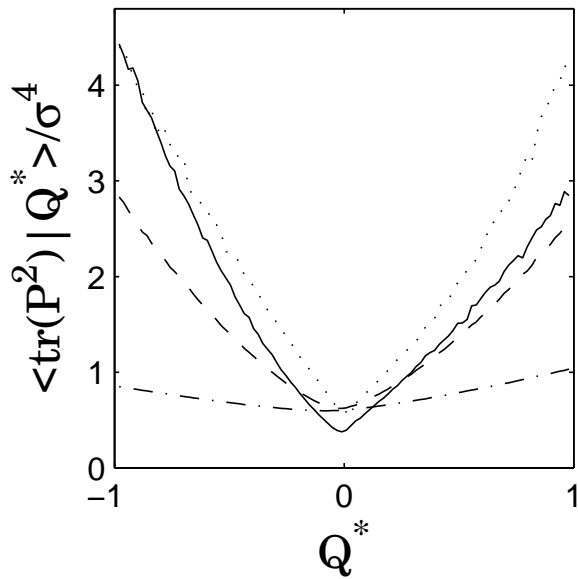,width=8cm}} \caption{Conditional expectation $\langle\mbox{tr}(\textbf{P}^2)|Q^*\rangle$ of 
pressure Hessian norm on $Q^*$: DNS (solid), Gaussian (dot-dashed), Multifractal (dashed) and MTLM (dotted) velocity fields. 
} \label{fig:CondPHonQ}
\end{figure}

As previously noticed in Ref. \cite{CheMen08}, current closures for the pressure Hessian \cite{ChePum99,CheMen06,GibHol07} 
are proportional  to the invariant $Q$. This is in particular the case for the closure provided in Ref. \cite{CheMen06}, namely 
\begin{equation}\label{eq:CMPH}
 \textbf{P} = -\frac{\textbf{C}_{\tau_\eta}^{-1}}{\mbox{tr}(\textbf{C}_{\tau_\eta}^{-1})}\mbox{tr}(\textbf{A}^2)\mbox{ ,}
\end{equation}
where $\textbf{C}_{\tau_\eta}$ is the statistically stationary Cauchy-Green tensor at the Kolmogorov time scale $\tau_\eta$ \cite{CheMen06,CheMen08} and $\mbox{tr}(\textbf{A}^2)=-2Q$. Indeed, it is  tempting to close the anisotropic part of \textbf{P} as a symmetric tensor proportional to $Q$, since the isotropic part is itself 
proportional to $Q$ as seen on the Poisson equation $\Delta p = 2Q$. It would imply in particular that the probability current 
$\vct{\mathcal W_p}$ vanishes on the $Q=0$ line. This is not observed on DNS (see Figs. \ref{fig:VectAmpRQ}(a) and (b)). 
To quantify more precisely the behavior of \textbf{P} in the neighborhood of vanishing $Q$, we proposed in Ref. \cite{CheMen08} 
to estimate the average pressure Hessian (square) norm $|\textbf{P}|^2 = \mbox{tr}(\textbf{P}^2)$ conditioned on the invariant $Q$. The 
corresponding conditional average $\langle\mbox{tr}(\textbf{P}^2)|Q\rangle$ was shown for fields obtained by DNS as well as by applying the closure (Eq. 
(\ref{eq:CMPH})). It was observed that for the DNS case, such conditional average at $Q=0$ does not vanish and furthermore it behaves nearly linearly with $Q$, whereas the closure Eq.  (\ref{eq:CMPH}) predicts a vanishing conditional average for $Q=0$ and a quadratic behavior with $Q$.

We display in Fig. \ref{fig:CondPHonQ} the conditional average $\langle\mbox{tr}(\textbf{P}^2)|Q\rangle$ as a function of $Q$ for 
the four different velocity fields: the DNS, Gaussian (Eq. (\ref{eq:GaussVect})), multifractal (Eq. (\ref{eq:MultiVel})) and MTLM
(Eq. (\ref{eq:MTLMVel})) velocity fields. As previously observed, for the DNS case (solid line), the conditional average does not vanish at $Q=0$ 
and behaves linearly with $Q$ in the neighborhood around $Q=0$. For all the three remaining synthetic velocity fields, the conditional average 
does not vanish at $Q=0$ and hence they perform better than the deterministic closure (Eq. \ref{eq:CMPH}) on this issue. For the Gaussian case however (dash-dotted line), the tails of the conditional average are not realistic, being far below the curves corresponding to the DNS case. Numerical simulations of the Gaussian fields at lower resolutions, i.e. $N=256^3$ or $N=512^3$  (data not shown), showed no difference with the $N=1024^3$ case. Interestingly, in this regard the multifractal field (dashed line) performed much better, exhibiting conditional average tails very close to the DNS result. Some discrepancy is found for negative $Q$s where, the tail 
in the DNS case has a steeper slope. The MTLM velocity field (dotted line) also performs well against DNS data, although its tails are found to be quite 
symmetric, at odds with DNS.

Overall, the behavior of the pressure Hessian obtained from the three synthetic velocity fields is reasonably satisfactory when compared 
against DNS data. The probability current is well reproduced in the $R<0$ region and the conditional average shown in  
Fig. \ref{fig:CondPHonQ} does not vanish for vanishing $Q$. Only the MTLM velocity field can reproduce additionally the void in probability observed in DNS over the $R^*>0$ region (explaining, or at least giving an interpretation of the lack of action of the pressure Hessian in the dissipation production dominated region $R^*>0$ remains, however,  an open problem). Furthermore, we have shown that synthetic velocity fields do predict the 
pressure Hessian square norm as being closer to linearly proportional to the invariant $Q$, rather than proportional to $Q^2$  as is the case in existing closures, in particular Eq. (\ref{eq:CMPH}). 

At this stage, one could reach the conclusion that some approximate surrogates of an actual turbulent field, even when obtained with the simplest Gaussian approximation, contain a better prediction of the behavior of the pressure Hessian, in connection with its dependence on $Q$, than the deterministic closure given by Eq. (\ref{eq:CMPH}). We will see in the following that the anisotropic part of the pressure Hessian can in fact be accurately closed by the local spatial variations of the invariant $Q$. 

\section{Locality of the pressure Hessian}

We have seen in the first part of this work that a simple Gaussian approximation, given by Eq. (\ref{eq:GaussVect}), 
or more sophisticated synthetic velocity fields, such as Eqs. (\ref{eq:MultiVel}) and (\ref{eq:MTLMVel}), can reproduce the motion of the 
probability current $\vct{\mathcal W}_p$ associated with the pressure Hessian in the $R^*<0$ region. Additionally, taking into account the 
spatial distribution of the velocity field also leads to a non-vanishing conditional mean pressure Hessian norm for $Q=0$. 
In this section we study to what degree spatial locality is important in determining these properties of the
anisotropic part of the pressure Hessian. 

The exact expression (\ref{eq:SOPH}) for the pressure Hessian is very useful since it allows interpreting its isotropic part 
as being local, whereas the anisotropic part is governed by  $\mbox{tr}(\textbf{A}^2)$, or equivalently $Q$ at different locations, i.e. it contains 
non-local contributions from the spatial variations of $Q$. In this section, we will work with an equivalent form of Eq. (\ref{eq:SOPH}) that underlines the role played by the Hessian \textbf{Q} of the invariant $Q$. Indeed, taking two spatial derivatives of the 
Poisson equation that  commute with the Laplacian, one obtains $\Delta \textbf{P} =2\textbf{Q}$, where $Q_{ij} = \frac{\partial^2 Q}{\partial x_i\partial x_j}$. 
It is then easily seen that a similar relation exists between the deviatoric parts of $\textbf{P}$ and $\textbf{Q}$, namely 
$\Delta \textbf{P}^d =2\textbf{Q}^d$, where the superscript $^d$ denotes the deviatoric part, i.e. for example 
$\textbf{P}^d = \textbf{P} - \frac{1}{3} \mbox{tr}(\textbf{P}) \textbf{I}$, with $\textbf{I}$ being the identity matrix. 
We finally reach a relation, equivalent to Eq. (\ref{eq:SOPH}), between $\textbf{P}^d$ 
and $\textbf{Q}^d$:
\begin{equation}\label{eq:PdToQd}
 \textbf{P}^d(\textbf{x}) = -\frac{1}{2\pi}\int \frac{1}{|\textbf{x}-\textbf{y}|}\textbf{Q}^d(\textbf{y}) d\textbf{y}\mbox{ .}
\end{equation}
Relation (\ref{eq:PdToQd}) is exact. In the following, we will truncate the integral present in Eq. (\ref{eq:PdToQd}) over a ball, centered at $\textbf{x}$, and of radius $\eta$, namely
\begin{equation}\label{eq:PdToQdTrunc}
 \textbf{P}^d(\textbf{x}) \approx -\frac{1}{2\pi}\int_{|\textbf{x}-\textbf{y}|\le \eta} \frac{1}{|\textbf{x}-\textbf{y}|}\textbf{Q}^d(\textbf{y}) d\textbf{y}\mbox{ .}
\end{equation}
It is easily seen that from Eq. (\ref{eq:PdToQdTrunc}) we recover Eq. (\ref{eq:PdToQd}) by taking $\eta\rightarrow +\infty$. We now make 
the strong assumption that $\eta$ is of order of the Kolmogorov length scale $\eta_K$. In this case, we can Taylor-expand 
the Hessian of $Q$ at the position $\textbf{y}$ around its value at the location $\textbf{x}$, take out $\textbf{Q}^d(\textbf{x})$ 
from the integral, and perform the remaining integration in spherical coordinates. We get an expression for the deviatoric part 
of the pressure Hessian
\begin{equation}\label{eq:LocalClos}
 \left(\frac{\partial^2 p}{\partial x_i\partial x_j}\right)^d \approx -\eta^ 2\left(\frac{\partial^2 Q}{\partial x_i\partial x_j}\right)^d\mbox{ .}
\end{equation}
This now expresses the anisotropic part of the pressure Hessian in terms of local properties of the invariant $Q$, although the latter's spatial derivatives
are needed.  These derivatives are unknown in the closure and Lagrangian models of e.g. \cite{CheMen06} and so this does not constitute a practical closure yet. 
To ensure that this expression yields the same norm as the true pressure Hessian, we define the ball's radius according to
\begin{equation}\label{eq:ChoiceEta}
 \eta^2 = \sqrt{\frac{\langle \mbox{tr}(\textbf{P}^2)\rangle}{\langle \mbox{tr}(\textbf{Q}^2)\rangle}}\mbox{ .}
\end{equation}
The expression (\ref{eq:LocalClos}) and the choice of the length scale $\eta$ by Eq. (\ref{eq:ChoiceEta}) are consistent only if (i) 
eigenvalues and eigenvectors of $\textbf{P}^d$ and $\textbf{Q}^d$ are well correlated and (ii) $\eta$ is indeed of the order of the Kolmogorov 
length scale $\eta_K$, since we assume that in the neighborhood of $\textbf{x}$, 
$\textbf{Q}^d(\textbf{y})\approx \textbf{Q}^d(\textbf{x})$.

Both Hessian tensors, for the pressure and for $Q$, are computed in Fourier space, for the periodic DNS flows. The expression
(\ref{eq:LocalClos}) requires the computation of the second derivative of $Q$, which is already the square of a first spatial 
derivative. This is a reason for our use of the highly resolved DNS under study, in which $dx/\eta_K\approx 1$ (see section \ref{DNSvf}). Using $\eta_K = (\nu^3/\langle \epsilon \rangle)^{1/4} = 0.0123$, we find that for the current DNS our choice of the length scale $\eta$ as Eq. (\ref{eq:ChoiceEta}) implies that $\eta=1.83~\eta_K$. Hence, $\eta$ is of the order of  $\eta_K$, as required. Let us stress that a well resolved DNS is  necessary since some noise can be introduced by the computation of third order derivatives, leading to a bad estimation of the  parameter $\eta$. More work is needed to clarify this point and to assess the dependence of $\eta$ on resolution effects and Reynolds numbers. This could be done, for example, performing specifically designed DNS aimed at quantifying accurately high-order velocity derivatives, as was proposed in Refs. \cite{SheChe93,Sch07}.

\begin{figure}[t]
\center{\epsfig{file=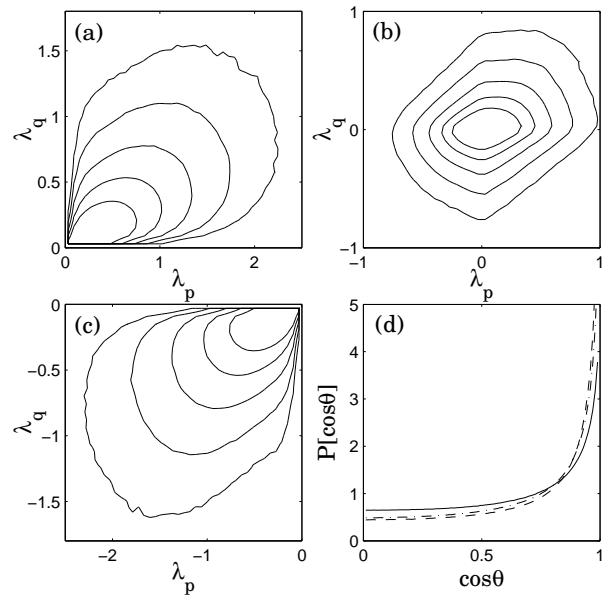,width=8cm}} \caption{In (a,b,c), we show the joint PDF of eigenvalues $\lambda_p$ of the pressure Hessian and eigenvalues $\lambda_q$ of the proposed local expression
(Eq. (\ref{eq:LocalClos})). Iso-probability lines correspond to $e^{-4},e^{-3},e^{-2},e^{-1},1$. In (d) is shown the PDF of the cosine of 
the angle between the eigenvectors of \textbf{P} and those of the local closure (Eq. (\ref{eq:LocalClos})): eigenvectors associated to the smallest (dashed), intermediate (solid) and biggest (dot-dashed) eigenvalues.} \label{fig:EigenPlotLev}
\end{figure}

We display in Fig. \ref{fig:EigenPlotLev} the joint probability densities of the three eigenvalues of the deviatoric part of the 
true pressure Hessian, (denoted by $\lambda_p$) and the corresponding eigenvalues of the expression $-\eta^2\textbf{Q}^d$ 
(denoted by $\lambda_q$). The smallest (Fig. \ref{fig:EigenPlotLev}(a)) and largest eigenvalues (Fig. \ref{fig:EigenPlotLev}(c)) 
are found to be well correlated with joint density contours are elongated along the perfect correlation line 
(i.e. $\lambda_p=\lambda_q$), and
correlation coefficients of $\rho = 0.714$ and  $\rho = 0.703$ respectively.

As far as the intermediate  eigenvalue is concerned, weaker correlation is found ($\rho = 0.275$), with isolines being close to 
circles. We can conclude that the smallest and 
largest eigenvalues are well correlated.  In Fig. \ref{fig:EigenPlotLev}(d) we display the probability of the cosine of the angle 
between the eigenvectors 
of the left and right terms of Eq. (\ref{eq:LocalClos}). It is found that in all the cases, corresponding to the three different 
eigenvalues, the maximum of probability is reached when the eigenvectors of $\textbf{P}^d$ and $-\eta^2\textbf{Q}^d$ are aligned. 
These results show then that $\textbf{P}^d$ and $-\eta^2\textbf{Q}^d$ are correlated both in amplitude and eigendirections. 

\begin{figure}
\center{\epsfig{file=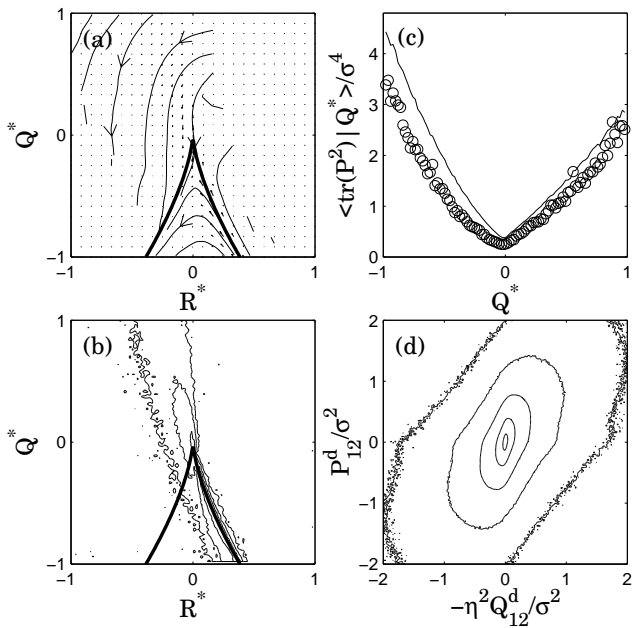,width=8.5cm}} \caption{(a) and (b): Probability current $\vct{\mathcal W_p}$ 
associated with pressure Hessian 
(Eq. \ref{eq:WP}) for the DNS velocity field when using the local closure (\ref{eq:LocalClos}. See caption of Fig. 
\ref{fig:VectAmpRQ} for further details. (c): Conditional expectation $\langle\mbox{tr}(\textbf{P}^2)|Q\rangle$ of 
pressure Hessian norm on $Q^*$: DNS (solid) and closure ($\circ$), in a similar fashion than in Fig. \ref{fig:CondPHonQ}. 
(d) Joint PDF of the components $P_{12}$ and $-\eta^2AQ_{12}$ in a logarithmic scale. Isolines correspond to probability 
$[10^{-3},10^{-2},10^{-1},1,10]$.} \label{fig:VectAmpHQ}
\end{figure}

We have seen that, in tensorial structure, the true pressure Hessian $\textbf{P}^d$ and the local expression in terms of velocity gradients $-\eta^2\textbf{Q}^d$ are quite
similar. We may now wonder if the local expression is able to reproduce the probability current associated to pressure effects as seen 
in Figs. \ref{fig:VectAmpRQ}(a-b). To this purpose, we show in Figs. \ref{fig:VectAmpHQ}(a-b) the probability current $\vct{\mathcal W_p}$ obtained 
from DNS when the true pressure Hessian is replaced by the local expression (\ref{eq:LocalClos}). We see in Fig. \ref{fig:VectAmpRQ}(a) that indeed the local expression reproduces the counteractive action of the pressure along the right tail of the Vieillefosse line. Furthermore, a void in probability is found in dissipation production dominated regions (i.e. $R^*>0$), 
as  can be clearly seen in Fig. \ref{fig:VectAmpRQ}(b). Finally, the local expression also reproduces some of the probability redistribution 
in the enstrophy production dominated region ($R^*<0$), as observed in DNS (Fig. \ref{fig:VectAmpRQ}(a-b)), although the direction of the probability flux  streamlines
for $R^*<0$ is seen to be more vertical than the left-wards directions seen in the DNS. Also, the  
streamlines are found more curved for the local expression than for  \textbf{P}  and the probability current amplitude 
$|\vct{\mathcal W_p}|$ is found to decrease faster at high values of $R$ and $Q$ than in the DNS case. This could be due to 
limitations of the localized expression in reproducing very high turbulent fluctuations

The overall behavior of the local expression in the $RQ$-plane is on the whole quite satisfactory when compared against DNS. Some differences appear: (i) the 
streamlines of the probability current are found more curved for the local closure than for \textbf{P}, very much in opposition 
to the streamlines imposed by the RE approximation (see Ref. \cite{CheMen08}), and (ii) the probability current amplitude 
$|\vct{\mathcal W_p}|$ is found to decrease faster at high values of $R$ and $Q$ than in the DNS case. This could be due to 
limitations of the localized expression in reproducing very high turbulent fluctuations. 
On the whole, however, the trends provided by the local expression Eq. (\ref{eq:LocalClos}) agree quite well compared to the DNS results in terms of the 
prbability fluxes in the $RQ$ plane.

We also show in Fig. \ref{fig:VectAmpHQ}(c) the conditional expectation of the pressure Hessian square norm, based on $Q$. For the sake of 
clarity, we  show again the conditional average as obtained in DNS (solid line) (it was already shown in Fig. \ref{fig:CondPHonQ}). 
We use open symbols ($\circ$) to show the conditional expectation obtained from DNS using the local approximation (\ref{eq:LocalClos}). 
It can been seen that the conditional expectation obtained from the local approximation is in very good agreement with the DNS, with some discrepancies appearing for 
negative $Q^*$. Interestingly, we  see that the conditional expectation does not vanish for   $Q=0$. 
This is consistent with former remarks made about the requirement that a realistic model of $\textbf{P}$ cannot be simply
proportional to $Q$.

Finally, to quantify the agreement between individual tensor elements,  we show in Fig. \ref{fig:VectAmpHQ}(d) the joint PDF 
of the component $P_{12}$ and the component $-\eta^2Q_{12}$. This plot demonstrates the good level of correlation between these 
two tensors since the joint PDF is clearly elongated along the perfect correlation line (i.e.  $P_{12} = -\eta^2Q_{12}$). The corresponding correlation
coefficient is $\rho = 0.55$. Thus, also this  statistical  test confirms the good agreement between the local approximation and the true pressure Hessian.  

\section{Conclusions and perspectives}

This article focuses on the statistical nature of the pressure Hessian that governs much of the Lagrangian dynamics of the velocity 
gradient tensor in turbulence. In a first part, we have seen that synthetic velocity fields reproduce many properties of the pressure Hessian 
as they are seen in DNS flows, such as the non-trivial behavior of the probability current, and the 
conditional expectation of the pressure Hessian norm on the invariant $Q$. Even the simplest Gaussian approximation for the velocity 
field (Eq. (\ref{eq:GaussVect})) can represent some non-trivial behaviors of the pressure that could not be predicted by the closures in terms 
of \textbf{A}  \cite{ChePum99,CheMen06,GibHol07}. Based on this observation and on an exact 
field description of the pressure Hessian by means of nonlocal integrals (Eqs. (\ref{eq:SOPH}) and (\ref{eq:PdToQd})), we formulate the hypothesis that considering only the integration over a ball of radius $\eta_K$, and neglecting other contributions, the deviatoric part of \textbf{P} could be expressed in terms of local properties of the 
velocity gradient tensor, but in terms of higher-order derivatives. Specifically, the spatially local approximation is not expressed 
in terms of $\textbf{A}$ but in terms of second-order derivatives of $Q$. This approximation was found to be highly correlated 
with the true pressure Hessian \textbf{P} when compared in DNS computations. These findings show that the main contribution to 
\textbf{P}(\textbf{x}) is contained in the local neighborhood around position \textbf{x}, in a ball centered at \textbf{x} and 
of radius  of the order of 
$\eta_K$. This raises the hope that  local closures involving a finite set of ordinary differential equations may still be possible, for studying the Lagrangian dynamics of the velocity gradient tensor.  To that end, it is still necessary to  express the Hessian of $Q$ in terms  of the local values of $\textbf{A}$. Only then would we have a full closure.

Let us finally remark that if a tractable transport equation for the pressure Hessian is difficult to get, 
the Lagrangian derivatives of $p$ and $\textbf{P}$ can be related. We get the following transport equation for the pressure Hessian
\begin{equation}\label{eq:TransPH}
\frac{d^{O}\textbf{P}}{dt} = - \frac{\partial p}{\partial x_k}\vct{\nabla}\vct{\nabla} u_k + 
\vct{\nabla}\vct{\nabla}\frac{dp}{dt} \mbox{ ,}
\end{equation}
where $d^O/dt$ stands for the upper convected time derivative or Oldroyd derivative, that relates the rate of change 
written in the coordinate system rotating and stretching with the fluid (see for example Ref. \cite{Hau02}), i.e.
\begin{equation}\label{eq:Oldroyd}
 \frac{d^{O}\textbf{P}}{dt} = \frac{d\textbf{P}}{dt} + \textbf{A}^\top \textbf{P} + \textbf{P}\textbf{A}\mbox{ .}
\end{equation}
Then, if we neglect the right-hand side of Eq. (\ref{eq:TransPH}) in the rate of change of the pressure Hessian, i.e. 
if we assume that the 
Oldroyd derivative (\ref{eq:Oldroyd}) vanishes, and we apply the recent fluid deformation (RFD) approximation 
(assuming $\textbf{A}$ independent on time), we get that for an early time
\begin{equation}
 \textbf{P}(\tau) = e^{-\tau \textbf{A}^\top}\textbf{P}(0)e^{-\tau \textbf{A}}\mbox{ .}
\end{equation}
This remark justifies the use of matrix exponentials as closures of the pressure Hessian \cite{CheMen06}, 
as it was also noted for the subgrid-scale stress tensor \cite{LiChe09}. If in addition we start from an initial 
isotropic pressure 
Hessian, and include the Poisson equation, we recover exactly the closure (\ref{eq:CMPH}) proposed in Ref. \cite{CheMen06}. 
This represents useful perspectives for future investigations on the Lagrangian dynamics of 
the velocity gradient tensor.

We thank G. Eyink and B. L\"uthi for fruitful discussions. We acknowledge the Leonardo Da Vinci Programme for support in the internship of 
F.T. at the ENS, and the CNRS for constant support. DNS and stochastic fields have been performed by using the local 
computing facilities (PSMN) at ENS Lyon 
under grant CPER-CIRA. This research was supported in part by the National Science Foundation 
under Grant No. NSF PHY05-51164 during the stay of L.C. at the KITP (Santa Barbara).
CM and HY are supported by NSF grant CDI-0941530.
C.R. acknowledges the support of CONICYT under Project Fondecyt 11080025.


\end{document}